\documentclass{article}
\usepackage{spconf,amsmath,graphicx,bbm,array,dcolumn,epsfig,amsmath,amsfonts,yhmath,bm,amssymb,psfrag,color,soul,enumerate,stackengine,cite}
\usepackage[dvipsnames]{xcolor}
\usepackage{tcolorbox}
\usepackage[font=footnotesize,skip=7pt]{caption}
\usepackage[font=small,skip=0pt]{subcaption}

\usepackage[lining,scaled=1.05]{ebgaramond}

\usepackage{tikz}

\definecolor{lgreen} {RGB}{180,210,100}
\definecolor{ngreen} {RGB}{98,158,31}
\definecolor{dgreen} {RGB}{78,138,21}
\definecolor{MLOWLSgreen} {RGB}{0,140,130}
\definecolor{SDPpurple} {RGB}{191,0,191}
\definecolor{lred}   {RGB}{220,0,0}
\definecolor{nred}   {RGB}{224,0,0}
\definecolor{bred}   {RGB}{200,20,20}
\definecolor{nblue}  {RGB}{28,130,185}
\definecolor{jblue}  {RGB}{20,50,100}
\newcommand*\circled[1]{\tikz[baseline=(char.base)]{
		\node[circle,draw,color=bred, opacity=0.75,inner sep=1pt] (char) {\footnotesize #1};}}

\newcommand {\myvec}[1] {{\mbox{\boldmath $#1$}}}
\newcommand {\mymat}[1]  {{\mbox{\boldmath $#1$}}}

\DeclareMathAlphabet      {\mathbfit}{OML}{cmm}{b}{it}

\AtBeginEnvironment{bmatrix}{\setlength{\arraycolsep}{4pt}}

\newcommand{\etal}{\textit{et al.\ }}

\newcommand {\A} {\mymat{A}}

\newcommand {\mLambda} {\mymat{\Lambda}}
\newcommand {\hmLambda} {\widehat{\mLambda}}

\newcommand {\tC} {\widetilde{C}}
\newcommand {\tmC} {\widetilde{\C}}
\newcommand {\C} {\mymat{C}}

\newcommand {\mPhi} {\mymat{\Phi}}
\newcommand {\mPsi} {\mymat{\Psi}}

\newcommand {\uphi} {\myvec{\phi}}
\newcommand {\upsi} {\myvec{\psi}}

\newcommand {\Ep} {\mymat{\mathcal{E}}}
\renewcommand {\H} {\mymat{H}}

\newcommand {\tpsi} {\widetilde{\psi}}
\newcommand {\tupsi} {\widetilde{\upsi}}

\renewcommand {\P} {\mymat{P}}
\newcommand {\tP} {\widetilde{\P}}

\newcommand {\R} {\mymat{R}}
\newcommand {\hR} {\widehat{\R}}
\newcommand {\I} {\mymat{I}}

\newcommand {\ua} {\myvec{a}}

\newcommand {\uepsilon} {\myvec{\epsilon}}

\newcommand {\uc} {\myvec{c}}

\newcommand {\uxi} {\myvec{\xi}}

\newcommand {\ualpha} {\myvec{\alpha}}
\newcommand {\uv} {\myvec{v}}

\newcommand {\uo} {\myvec{0}}
\newcommand {\us} {\myvec{s}}

\newcommand {\ux} {\myvec{x}}
\newcommand {\ur} {\myvec{r}}
\newcommand {\urho} {\myvec{\rho}}
\newcommand {\uiota} {\myvec{\iota}}
\newcommand {\uy} {\myvec{y}}
\newcommand {\uw} {\myvec{w}}

\newcommand {\utheta} {\myvec{\theta}}

\newcommand {\uon} {\myvec{1}}

\newcommand {\ep} {\mathcal{E}}
\newcommand {\uvarep} {\myvec{\varepsilon}}
\newcommand {\ueta} {\myvec{\eta}}

\newcommand {\Rset} {\mathbb{R}}
\newcommand {\Cset} {\mathbb{C}}

\newcommand {\Eset} {\mathbb{E}}
\newcommand {\Nset} {\mathbb{N}}

\newcommand {\Diag} {\text{\normalfont Diag}}

\newcommand {\tps} {\tiny \rm{T}}
\newcommand {\her} {\tiny \rm{H}}

\title{NON-ITERATIVE BLIND CALIBRATION OF NESTED ARRAYS WITH ASYMPTOTICALLY OPTIMAL WEIGHTING}

\name{Amir Weiss$^{\star}$ and Arie Yeredor$^\dagger$}

\address{
\begin{tabular}{cc}
$^{\star}$Dept. of Computer Science and Applied Mathematics & $^\dagger$School of Electrical Engineering\\
Weizmann Institute of Science & Tel-Aviv University\\
amir.weiss@weizmann.ac.il & arie@eng.tau.ac.il
\end{tabular}
}
\begin{document}
\ninept
\maketitle
\setlength{\abovedisplayskip}{5pt}
\setlength{\belowdisplayskip}{5pt}

\begin{abstract}
\small{Blind calibration of sensors arrays (without using calibration signals) is an important, yet challenging problem in array processing. While many methods have been proposed for ``classical" array structures, such as uniform linear arrays, not as many are found in the context of the more ``modern" sparse arrays. In this paper, we present a novel blind calibration method for $2$-level nested arrays. Specifically, and despite recent contradicting claims in the literature, we show that the Least-Squares (LS) approach can in fact be used for this purpose with such arrays. Moreover, the LS approach gives rise to optimally-weighted LS joint estimation of the sensors' gains and phases offsets, which leads to more accurate calibration, and in turn, to higher accuracy in subsequent estimation tasks (e.g., direction-of-arrival). Our method, which can be extended to $K$-level arrays ($K>2$), is superior to the current state of the art both in terms of accuracy and computational efficiency, as we demonstrate in simulation.}
\end{abstract}

\vspace{-0.3cm}
\section{Introduction}\label{sec:intro}
\vspace{-0.2cm}
Non-uniform linear sparse arrays, such as nested \cite{pal2010nested} and coprime \cite{liu2016coprime} arrays, are at the heart of the recently emerging and promising field of sparse array signal processing \cite{boudaher2015sparsity,liu2017one,gupta2018design,nannuru2018sparse,cohen2018optimized}. With relatively simple geometries, leading to the notion of virtual sensors, these sparse arrays enable the detection of $\mathcal{O}(N^2)$ sources with only $\mathcal{O}(N)$ physical sensors \cite{liu2017cramer}.

While the related theory continues to rapidly evolve, including, for instance, adapted Direction-of-Arrival (DOA) estimation methods \cite{zhou2018direction} and related performance bounds \cite{wang2016coarrays}, the problem of \emph{blind calibration} (e.g., \cite{weiss2019blindcalibration}) has seen relatively less treatment in the literature thus far \cite{han2015calibrating,ramamohan2019blindbroad,ramamohan2019blind,yang2020calibrating}. Contrary to array calibration using a known, user-controlled source, blind calibration, in which even the number of received sources is (possibly) unknown, is typically a more desirable, yet a more challenging task.

In this paper, we address the problem of blind calibration of nested arrays comprised of two (collinear) Uniform Linear Arrays (ULAs). In particular, we derive closed-form expressions of Least Squares (LS) estimates, with approximate optimal weighting, of the unknown gain and phase offsets. Unlike previous methods, such weighting enables to exploit latent characteristics of the joint statistical behavior of errors in the estimated covariance.

Our main contributions in this work are the following. First, we show that the LS-based approach is in fact applicable to blind calibration of nested arrays, despite recent contradicting claims in the literature \cite{han2015calibrating}. We then derive Optimally-Weighted LS (OWLS) estimates for \emph{joint} estimation of the unknown sensors' gains and phases, exploiting cross-correlations between estimation errors in the respective gains and phases equations. Finally, we demonstrate that our solution is not only more accurate than the current state of the art \cite{ramamohan2019blind} by Ramamohan \textit{et al.}, but is also more computationally efficient by orders of magnitude.

%\vspace{-0.25cm}
%\subsection{Notations}\label{subsec:notations}
%\vspace{-0.15cm}
%The superscripts $(\cdot)^{\tps}$, $(\cdot)^*$, $(\cdot)^{\her}$ and $(\cdot)^{-1}$ denote the transpose, complex conjugate, conjugate transpose and inverse operators, resp. We denote $\I_{N}$ as the $N\hspace{-0.03cm}\times\hspace{-0.03cm}N$ identity matrix and $\uo_{N}\hspace{-0.02cm}\in\hspace{-0.02cm}\Rset^{N\times 1}$ as the all-zeros vector. $\Eset[\cdot]$ denotes expectation, $\text{vec}(\cdot)$ concatenates all the columns of its matrix argument in a column vector, $\Diag(\cdot)$ forms an $N\hspace{-0.03cm}\times\hspace{-0.03cm}N$ diagonal matrix from its $N$-dimensional vector argument, and $[N]$ denotes the set of integers $\{1,\ldots,N\}$. $\Re\{\cdot\}$ and $\Im\{\cdot\}$ denote the real and imaginary parts (resp.) of their complex-valued argument, and $\jmath$ denotes $\sqrt{-1}$;

\vspace{-0.22cm}
\section{Problem Formulation}\label{sec:problem}
\vspace{-0.22cm}
Consider a general $2$-level linear nested array with $N_1$ and $N_2$ sensors in its first and second levels, resp., namely, with a total of $N\triangleq N_1+N_2$ sensors. The first and second level ULAs have an inter-element spacing of $d$ and $\bar{d}\triangleq L\cdot d$, resp., where\footnote{Usually, $L=N_1+1$, as originally proposed by Pal and Vaidyanathan \cite{pal2010nested}.} $L\in\Nset$. We assume that the $N$ sensors have \textit{unknown} deterministic gain and phase offsets, which we denote as $\upsi\in\Rset_+^{N \times 1}$ and $\uphi\in[-\pi,\pi)^{N \times 1}$, resp. ($\psi_n$ and $\phi_n$ are the unknown gain and phase offsets of the $n$-th sensor, resp.). Further, we assume the presence of an unknown number $M$ of unknown, ``far-field" narrowband sources, centered around some common carrier frequency with wavelength $\lambda$.

The vector of sampled, baseband-converted signals from all $N$ sensors is given by
\begin{equation}\label{modelequation}
\ur(t)\hspace{-0.05cm}=\hspace{-0.05cm}\mPsi\mPhi\big(\A(\ualpha)\us(t)+\uv(t)\big)\triangleq\mPsi\mPhi\ux(t)\in\Cset^{N\times1}, \forall t\in[T],
\end{equation}
where $[N]$ denotes the set of integers $\{1,\ldots,N\}$ for any $N\in\Nset$, and
\begin{enumerate}[(i)]
	\itemsep0.05em 
	\item $\mPsi\triangleq\Diag(\upsi)\in\Rset_+^{N\times N}$, $\mPhi\triangleq\Diag\left(e^{\jmath\uphi}\right)\in\Cset^{N\times N}$;
	\item $\us(t)\triangleq\left[s_1(t)\;\cdots\;s_M(t)\right]^{\tps}\in\Cset^{M\times1}$ denotes the sources, originating from unknown angles $\ualpha\triangleq\left[\alpha_1\;\cdots\;\alpha_M\right]^{\tps}\in\Rset^{M\times1}$;
	\item $\A(\ualpha)\triangleq\left[\ua(\alpha_1)\;\cdots\;\ua(\alpha_M) \right]\in\Cset^{N\times M}$ denotes the nominal array manifold matrix, with the steering vectors $\ua(\alpha_m)$ as its columns, defined elementwise as $a_n(\alpha_m)\triangleq e^{\jmath \frac{2\pi}{\lambda}d\cdot i_n\cos(\alpha_m)}$ for all $n\in[N], m\in[M]$, 
%	\begin{equation*}
%		a_n(\alpha_m)\triangleq e^{\jmath k_{\lambda}di[n]\cos(\alpha_m)}, \forall n\in[N], \forall m\in[M],
%	\end{equation*}
	with the indexing function
	\begin{equation*}\label{indexfunctiondef}
		i_n\triangleq\begin{cases}
		n-1, & n\in[N_1]\\
		N_1 + (n-1-N_1)L, & (n-N_1)\in[N_2]
		\end{cases};
	\end{equation*}
	\item $\uv(t)\in\Cset^{N\times1}$ denotes additive ambient noise or ``interfering" signals, modeled as spatially and temporally independent, identically distributed (i.i.d.) zero-mean circular Complex Normal (CN) \cite{loesch2013cramer} with a covariance matrix $\R_v\triangleq\Eset\left[\uv(t)\uv(t)^{\her}\right]=\sigma_v^2\I_N$, where $\sigma_v^2$ is unknown; and
	\item $\ux(t)$ denotes the signal that would have been received in the absence of gain or phase offsets, namely with $\mPsi=\mPhi=\I_N$.
\end{enumerate}

We also assume that the sources are mutually uncorrelated. Particularly, $\us(t)$ is modeled as a (temporally) i.i.d.\ zero-mean circular CN vector process with an unknown diagonal covariance matrix $\R^s\triangleq\Eset\left[\us(t)\us(t)^{\her}\right]$. However, we note that the CN assumption can be relaxed while retaining most of the results obtained throughout this paper. Additionally, we assume $\us(t)$ and $\uv(t)$ are statistically independent. As a consequence,
\begin{equation}\label{CN_samples}
	\ur(t)\sim \mathcal{CN}\left(\uo_N,\R\right), \forall t\in[T],
\end{equation}
where
\begin{align}
		\hspace{-0.075cm}\R&\triangleq\Eset\left[\ur(t)\ur(t)^{\her}\right]=\mPsi\mPhi\C\mPhi^*\mPsi\in\Cset^{N\times N},\label{covariance_of_r}\\
		\hspace{-0.075cm}\C\hspace{-0.05cm}&\triangleq\hspace{-0.05cm}\Eset\left[\ux(t)\ux(t)^{\her}\right]\hspace{-0.05cm}=\hspace{-0.05cm}\A(\ualpha)\R^s\A^{\her}(\ualpha)+\sigma_v^2\I_N\in\Cset^{N\times N},\label{covariance_C}
\end{align}
and we have used $\mPsi^{\her}=\mPsi$ and $\mPhi^{\her}=\mPhi^*$.

The problem at hand can now be formulated as follows. \textit{Given the measurements $\left\{\ur(t)\right\}_{t=1}^{T}$ whose (identical) distribution is prescribed by \eqref{CN_samples}, estimate the unknown deterministic parameters $\left\{\upsi,\uphi\right\}$.}

Before proceeding to the solution approach, we also note that our proposed method can be modified and be applied to a more general signal model,
\begin{equation}\label{extendedmodel}
 \ur_w(t)=\ur(t)+\uw(t),
\end{equation}
where $\uw(t)$ denotes a spatially and temporally white additive noise vector, statistically independent of $\ur(t)$, which is unaffected by the gain and phase offsets $\upsi, \uphi$. This possible additional noise term usually accounts for internal (e.g., thermal) receiver noise. Details regarding the required modifications can be found in \cite{weiss2020asymptotically}, Subsection IV.

%\tcbset{colframe=gray!95!blue,size=small,width=0.49\textwidth,arc=2.1mm,outer arc=1mm}
%\begin{tcolorbox}[upperbox=visible,colback=white]
%{\textit{Given the statistically independent measurements $\left\{\ur[t]\right\}_{t=1}^{T}$ whose (identical) distribution is prescribed by \eqref{CN_samples}, estimate the unknown (deterministic) parameters $\left\{\upsi,\uphi\right\}$.}}
%\end{tcolorbox}

\vspace{-0.35cm}
\section{Blind Calibration by Joint OWLS}\label{sec:blindcalibrationOWLS}
\vspace{-0.2cm}
First, observe that, given an upper bound on the number of sources $M$, an optimal solution to our problem would be obtained by joint Maximum Likelihood (ML) estimation of $\upsi$ and $\uphi$, together with the nuisance parameters $\sigma_v^2, M, \ualpha$ and the diagonal elements of $\R^s$, which yields efficient estimates (\hspace{1sp}\cite{ra1922mathematical}) of $\upsi, \uphi$. However, since:
\begin{enumerate}
	\itemsep0.025em
\item[\circled{1}] $M$ is unknown, which necessitates solving a preliminary model order selection problem (e.g., \cite{han2013improved,garg2020source,weiss2019blind});
\vspace{-0.05cm}\item[\circled{2}] Given $M$, the derivation of the likelihood equations for this model is rather involved, which, at any rate, leads to a non-convex, high dimensional optimization problem; and
\vspace{-0.05cm}\item[\circled{3}] The sufficient statistic in this model is the sample covariance matrix of the data $\hR\triangleq\frac{1}{T}\sum_{t=1}^{T}{\ur(t)\ur(t)^{\her}}\in\Cset^{M\times M}$,
\end{enumerate}\vspace{-0.05cm}
we resort to OWLS estimation of $\upsi, \uphi$ based only on $\hR$. This approach leads to computationally simple estimates, obtained by solving a linear system of equations, and enables to harness much (if not all) of the information encapsulated in the second-order statistics.
\vspace{-0.4cm}
\subsection{The Partially-Toeplitz Structure of the Covariance Matrix}\label{subsec:Toeplitzstructe}
\vspace{-0.1cm}
Due to the specific structure of a nested array, the covariance matrix $\C$ in \eqref{covariance_C} has a special structure, which, through the relation to $\R$ in \eqref{covariance_of_r}, gives rise to a linear system of equations in a concise set of unknowns when applying an element-wise $\log$ transformation to $\R$. This subsection focuses on establishing these linear relations, which, to the best of our knowledge, are presented in the literature for the first time for nested arrays in here, and serve as the primary contribution of this paper. Subsection \ref{subsec:OWLSestimation} is then devoted to the approximately optimal extraction (estimation) of $\upsi$ and $\uphi$ from these linear relations.

We begin by recalling the well-known fact that the array manifold matrix of a ULA is a Vandermonde matrix (e.g., \cite{petersen2008matrix}). Thus, when all the signals involved are uncorrelated, the covariance matrix of a (perfectly calibrated) ULA is a Toeplitz matrix (e.g., \cite{gray2006toeplitz}). Since a $2$-level linear nested array is a concatenation of two ULAs, its covariance matrix $\C$ contains two ``overlapping" Toeplitz matrices on its block diagonal. Specifically, the matrices $\C^{(1)}\in\Cset^{(N_1+1)\times(N_1+1)}$ and $\C^{(2)}\in\Cset^{N_2\times N_2}$, defined by
\begin{align*}\label{Toeplitzblocksdef}
C^{(1)}_{ij}&\triangleq C_{ij}\triangleq c^{(1)}_{|i-j|+1}, \forall i,j\in[N_1+1],\\
C^{(2)}_{ij}&\triangleq C_{(i+N_1)(j+N_1)}\triangleq c^{(2)}_{|i-j|+1}, \forall i,j\in[N_2],
\end{align*}
are both Hermitian, positive-definite Toeplitz matrices, where $\uc^{(1)}\in\Cset^{(N_1+1)\times 1}$ and $\uc^{(2)}\in\Cset^{N_2\times 1}$ are their associated underlying ``Toeplitz-generating" vectors, resp. Note further that, by definition,
\begin{equation}\label{commonelement}
C^{(1)}_{(N_1+1)(N_1+1)}\equiv C^{(2)}_{11}\in\Rset_+ \; \Longrightarrow \; c^{(1)}_{1}\equiv c^{(2)}_{1}\in\Rset_+,
\end{equation}
therefore these matrices share (/``overlap" in) a single element. This can be intuitively explained by the fact that the $(N_1+1)$-th sensor can be regarded either as the last element of the $1^{\text{st}}$ $(N_1+1)$-element ULA, or as the $1^{\text{st}}$ element of the $2^{\text{nd}}$ $N_2$-element ULA. Further, let us denote the upper right block of $\C$, which corresponds to the remaining cross-covariance elements between signals received by sensors from the first and second levels, as $\tmC\in\Cset^{N_1\times(N_2-1)}$, with
\begin{equation*}
\tC_{ij}\triangleq C_{i(j+N_1+1)}, \forall i\in[N_1], \forall j\in[N_2-1].
\end{equation*}

Based on \eqref{covariance_of_r}, using $R_{ij}=C_{ij}\psi_i\psi_j\cdot e^{\jmath\left(\phi_i-\phi_j\right)}$, we have for
\begin{gather}\label{Log_equation}
\log\left(R_{ij}\right)\triangleq\mu_{ij}+\jmath\cdot\nu_{ij},
\end{gather}
the following relations: (for \eqref{imag_log_eq}, we assume $\left|\phi_{n}\right|\ll\pi,\forall n\in[N]$)
\begin{align}
\mu_{ij}&=\Re\left\{\log\left(R_{ij}\right)\right\}=\Re\left\{\log\left(C_{ij}\right)\right\}+\log(\psi_i)+\log(\psi_j),\label{real_log_eq}\\
\nu_{ij}&=\Im\left\{\log\left(R_{ij}\right)\right\}=\Im\left\{\log\left(C_{ij}\right)\right\}+\phi_i-\phi_j,\label{imag_log_eq}
\end{align}
%\begin{equation}\label{real_log_eq}
%\mu_{ij}=\Re\left\{\log\left(R_{ij}\right)\right\}=\Re\left\{\log\left(C_{ij}\right)\right\}+\log(\psi_i)+\log(\psi_j),
%\end{equation}
%\begin{equation}\label{imag_log_eq}
%\nu_{ij}=\Im\left\{\log\left(R_{ij}\right)\right\}=\Im\left\{\log\left(C_{ij}\right)\right\}+\phi_i-\phi_j,
%\end{equation}
for any pair of indices $i\leq j\in[N]$. More specifically, using the Toeplitz structure of $\C^{(1)}$ and $\C^{(2)}$, and the definitions
\begin{equation*}\label{defofrhoandiota}
\begin{aligned}
&\hspace{-0.11cm}\urho^{(1)}\hspace{-0.03cm}\triangleq\hspace{-0.03cm}\Re\{\log(\uc^{(1)})\},\; \urho^{(2)}\hspace{-0.03cm}\triangleq\hspace{-0.03cm}\Re\{\log(\uc^{(2)})\},\; \tP\hspace{-0.03cm}\triangleq\hspace{-0.03cm}\Re\{\log(\tmC)\},\\
&\hspace{-0.05cm}\uiota^{(1)}\hspace{-0.03cm}\triangleq\hspace{-0.03cm}\Im\{\log(\uc^{(1)})\},\;\,\hspace{0.01cm} \uiota^{(2)}\hspace{-0.03cm}\triangleq\hspace{-0.03cm}\Im\{\log(\uc^{(2)})\},\;\;\hspace{0.01cm}  \widetilde{\I}\hspace{-0.03cm}\triangleq\hspace{-0.03cm}\Im\{\log(\tmC)\},
\end{aligned}
\end{equation*}
as well as $\tpsi_n\hspace{-0.05cm}\triangleq\hspace{-0.05cm}\log(\psi_n)$ (where $\log(\cdot)$ operates elementwise), we have
\begin{align}\label{real_log_eq_specific}
\mu_{ij}=&\tpsi_i+\tpsi_j+\begin{cases}
\rho^{(1)}_{|i-j|+1}, & i,j\in[N_1+1]\\
\rho^{(2)}_{|i-j|+1}, & (i-N_1),(j-N_1)\in[N_2]\\
\widetilde{P}_{ij}, &  i\in[N_1], N_1+2\leq j\leq N\\
\end{cases},
\end{align}\vspace*{-0.25cm}
\begin{align}\label{imag_log_eq_specific}
\nu_{ij}=&\phi_i-\phi_j+\begin{cases}
\iota^{(1)}_{|i-j|+1}, & i,j\in[N_1+1]\\
\iota^{(2)}_{|i-j|+1}, & (i-N_1),(j-N_1)\in[N_2]\\
\widetilde{I}_{ij}, &  i\in[N_1], N_1+2\leq j\leq N\\
\end{cases}.
\end{align}

It is readily seen from \eqref{real_log_eq_specific}--\eqref{imag_log_eq_specific} that, collecting all the non-redundant\footnote{Due to the conjugate symmetry $\log(\R)=\log(\R^{\her})=\log(\R)^{\her}.$} terms $\{\mu_{ij},\nu_{ij}\}$ into a vector $\uy\in\Rset^{N^2\times 1}$, namely $0.5N(N+1)$ values of $\mu_{ij}$ for $i,j\in[N]$ with $i\leq j$, and $0.5N(N-1)$ values of $\nu_{ij}$ for $i,j\in[N]$ with $i<j$, one can construct a system of \emph{linear} equations
\begin{equation}\label{exactrelationyandtheta}
\uy=\H\utheta,
\end{equation}
where $\utheta \triangleq \left[\tupsi^{\tps} \, \uphi^{\tps} \, \utheta_{\textrm{\scriptsize nuisance}}^{\tps} \right]^{\tps}\in\Rset^{K_{\theta}\times 1}$
%\begin{equation*}\label{vec_of_unknowns}
%\utheta \triangleq \left[\tupsi^{\tps} \, \uphi^{\tps} \, \utheta_{\textrm{\scriptsize nuisance}}^{\tps} \right]^{\tps}\in\Rset^{K_{\theta}\times 1},
%\end{equation*}
is the the vector of unknowns, $K_{\theta}\triangleq4N+2N_1(N_2-1)+2$,
\begin{equation*}\label{vec_of_unknowns_nuisance}
\utheta_{\textrm{\scriptsize nuisance}}\hspace{-0.005cm}\triangleq\hspace{-0.005cm}\left[{\urho^{(1)}}^{\tps} \, {\uiota^{(1)}}^{\tps} \, {\urho^{(2)}}^{\tps} \, {\uiota^{(2)}}^{\tps} \, \textrm{vec}\left(\tP\right)^{\tps} \, \textrm{vec}\left(\widetilde{\I}\right)^{\tps} \right]^{\tps},
\end{equation*}
and $\H$ is the matrix of coefficients determined by \eqref{real_log_eq_specific}--\eqref{imag_log_eq_specific}. Due to space limitations, and considering it is rather technical and straightforward, we omit the explicit definition of $\H$ (see \cite{weiss2020asymptotically}, Appendix A, for similar principles regarding the construction of $\H$). 

Assuming $K_{\theta}\leq N^2$, although \eqref{exactrelationyandtheta} has $N^2$ equations and only $K_{\theta}$ unknowns, it can be verified that $\H$ is not full-rank. Indeed, from \eqref{commonelement} we have the degeneracies $\rho_1^{(2)}=\rho_1^{(1)}$, as well as $\iota_1^{(1)}=\iota_1^{(2)}=0$. Moreover, similarly as with ULAs, identifiability of $\upsi,\uphi$ in this blind scenario requires one gain and two phase references \cite{astely1999spatial}. Hence, w.l.o.g.\ we may set $\tpsi_1=\phi_1=\phi_2=0$. We can now eliminate these six parameters from $\utheta$, together with the six corresponding columns of $\H$, maintaining \eqref{exactrelationyandtheta} with the newly defined $\utheta\in\Rset^{K_{\theta}\times 1}$ and $\H\in\Rset^{N^2\times K_{\theta}}$, only now $K_{\theta}=4N+2N_1(N_2-1)-4$.

\begin{figure}[t]
	\begin{tikzpicture}
	\fill[white] (-1/1.5, 0) circle (0.2) node {\color{black}(\hspace*{0.5pt}a\hspace*{0.5pt})};
	\fill[cyan!40!gray] (0/1.5, 0) circle (0.2) node {\color{white}1};
	\fill[cyan!40!gray] (1/1.5, 0) circle (0.2) node[label=above:{\color{black}First level}] {\color{white}2};
	\fill[cyan!40!gray] (2/1.5, 0) circle (0.2) node {\color{white}3};
	\draw [dotted, thick] (2.5/1.5,-0.225) -- (2.5/1.5,0.225);
	\fill[cyan!40!gray] (3/1.5, 0) circle (0.2) node {\color{white}4};
	\filldraw[color=cyan!40!gray, fill=white] (4/1.5, 0) circle (0.2) node {\color{cyan!40!gray}5};
	\filldraw[color=cyan!40!gray, fill=white] (5/1.5, 0) circle (0.2) node {\color{cyan!40!gray}6};
	\filldraw[color=cyan!40!gray, fill=white] (6/1.5, 0) circle (0.2) node {\color{cyan!40!gray}7};
	\fill[cyan!40!gray] (7/1.5, 0) circle (0.2) node[label=above:{\color{black}Second level}] {\color{white}8};
	\filldraw[color=cyan!40!gray, fill=white] (8/1.5, 0) circle (0.2) node {\color{cyan!40!gray}9};
	\filldraw[color=cyan!40!gray, fill=white] (9/1.5, 0) circle (0.2) node {\color{cyan!40!gray}10};
	\filldraw[color=cyan!40!gray, fill=white] (10/1.5, 0) circle (0.2) node {\color{cyan!40!gray}11};
	\fill[cyan!40!gray] (11/1.5, 0) circle (0.2) node {\color{white}12};
	\end{tikzpicture}
	\\
	\begin{tikzpicture}
	\fill[white] (-1/1.5, 0) circle (0.2) node {\color{black}(b)};
	\fill[cyan!40!gray] (0/1.5, 0) circle (0.2) node {\color{white}1};
	\fill[cyan!40!gray] (1/1.5, 0) circle (0.2) node {\color{white}2};
	\fill[cyan!40!gray] (2/1.5, 0) circle (0.2) node {\color{white}3};
	\draw [dotted, thick] (2.5/1.5,-0.225) -- (2.5/1.5,0.225);
	\fill[cyan!40!gray] (3/1.5, 0) circle (0.2) node {\color{white}4};
	\filldraw[color=cyan!40!gray, fill=white] (4/1.5, 0) circle (0.2) node {\color{cyan!40!gray}5};
	\filldraw[color=cyan!40!gray, fill=white] (5/1.5, 0) circle (0.2) node {\color{cyan!40!gray}6};
	\fill[cyan!40!gray] (6/1.5, 0) circle (0.2) node {\color{white}7};
	\filldraw[color=cyan!40!gray, fill=white] (7/1.5, 0) circle (0.2) node {\color{cyan!40!gray}8};
	\filldraw[color=cyan!40!gray, fill=white] (8/1.5, 0) circle (0.2) node {\color{cyan!40!gray}9};
	\fill[cyan!40!gray] (9/1.5, 0) circle (0.2) node {\color{white}10};
	\end{tikzpicture}
	\caption{$2$-level nested array with $N_1=N_2=3$ and $d=1$. (a) Conventional design: $L=N_1+1$, does not enable LS-based blind calibration. (b) Proposed design: $L=N_1$, enables LS-based blind calibration. Filled bullets represent sensors, while empty bullets represent spaces (absence of a sensors).}
	\label{fig:nestedarraydiagram}\vspace*{-0.5cm}
\end{figure}
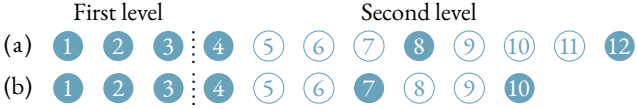

Yet, even after eliminating these degenerate variables, the rank of $\H$ is equal to $K_{\theta}-1$. More precisely, since the gain-related equations stemming from \eqref{real_log_eq_specific} are decoupled from the phase-related equations stemming from \eqref{imag_log_eq_specific}, it can be shown that (at least) one additional equation is required only for the unique determination of the phases (and not for the gains). Indeed, for a conventional $2$-level nested array with $L=N_1+1$, illustrated in Fig.\ \ref{fig:nestedarraydiagram}a for $N_1=3$, Han \etal\ \cite{han2015calibrating} already pointed out that ``there are not enough equations, as in the ULA case, to estimate the phase errors". Nevertheless, the claim that ``the least squares method becomes inapplicable" (\hspace{1sp}\cite{han2015calibrating}, Subsection IV-B) is not entirely accurate, as shown next.

Consider the case where $N_1=N_2=L$, as illustrated in Fig.\ \ref{fig:nestedarraydiagram}b for $N_1=3$. Contrary to the conventional choice $L=N_1+1$, which does \emph{not} allow for LS-based blind calibration of the array, with $L=N_1$ the situation is different, and LS-based blind calibration is in fact possible. To see this, observe that if we eliminate the $2^{\text{nd}}$ to $N_1$-th elements of the array, we are left with a ULA with $\bar{d}=N_1\cdot d$. This means that if we eliminate the respective ($2^{\text{nd}}$ to $N_1$-th) rows and columns of $\C$, we are left with a Toeplitz matrix, whose lower-right $N_2\times N_2$ block is $\C^{(2)}$. It follows that all elements of $\C^{(2)}$ are replicated versions of some of the elements of the $1^{\text{st}}$ row of $\C$, thus $\urho^{(2)}$ and $\uiota^{(2)}$ can be eliminated from $\utheta$, together with their corresponding columns in $\H$ (modifying columns corresponding to the other elements of $\utheta$ accordingly), which is now full-rank, hence \eqref{exactrelationyandtheta} has a unique solution. Nevertheless, it is still possible to detect $\mathcal{O}(N^2)$ sources using only $\mathcal{O}(N)$ sensors with a calibrated array. Indeed, it is easily verified that the maximum central contiguous ULA segment (\hspace{1sp}\cite{liu2016coprime}, Section II, definition 2) is of cardinality $2N_1^2+1=\mathcal{O}(N^2)$, which directly determines the maximum number of detectable sources (see \cite{liu2016coprime}, and reference therein). Moreover, the degradation in this respect relative to what can be achieved by $L=N_1+1$ is quite marginal. Thus, subject to this marginal cost, we derived a system of linear equations with a unique solution, serving as the underlying ``transformed" model for estimation of $\utheta$, which includes $\upsi$ and $\uphi$, based only on $\hR$.

We further note that if the number of detectable sources is of greater importance than the \emph{fully} blind calibration ability, the results of Subsection \ref{subsec:OWLSestimation} can also be used with $L=N_1+1$ by adding (only) one extra (third) phase reference to resolve the rank deficiency of $\H$.

\vspace{-0.2cm}
\subsection{Joint OWLS of the Sensors' Gains and Phases Offstes}\label{subsec:OWLSestimation}
\vspace{-0.2cm}
The transformed linear model \eqref{exactrelationyandtheta} enables estimation of $\upsi$ (via $\tupsi$) and $\uphi$, both extracted from $\utheta$, by replacing the unknown elements of $\uy$, namely $\{\mu_{ij},\nu_{ij}\}$, with their estimates $\{\widehat{\mu}_{ij},\widehat{\nu}_{ij}\}$, extracted from $\hR$. Indeed, theoretically, $\hR$ can be made arbitrarily close to $\R$ by increasing the sample size $T$. However, in practice, the available sample size is always limited and is oftentimes fixed. Therefore, rather than relying on the coarse approximation $\hR\approx\R$, which leads to the coarse, sub-optimal ordinary LS estimate, we propose a more refined analysis, which takes into account the estimation errors in $\hR$ and exploits (some of) their approximated statistical properties for obtaining a more accurate estimate of $\utheta$.

%\begin{figure}[t]
%	\centering
%	\includegraphics[width=0.43\textwidth]{plot_weight_function}
%	\vspace{-0.1cm}
%	\caption{The weight function $w[n]$ of the difference co-array, as defined in \cite{pal2010nested}, Section II. With $L=N_1$, the maximum central contiguous ULA segment (red) is still $\mathcal{O}(N^2)$. The green and pink circles mark the difference $N_1$: with $L=N_1$ (pink), $w[N_1]=3>1$, namely there exist a common phase ``linking" the two ULAs \eqref{extraequation4phase}, one which is not available with $L=N_1+1$ (green), $w[N_1]=1$.}
%	\label{fig:weightfunction}\vspace*{-0.6cm}
%\end{figure}
Formally, for any finite sample size $T$, we have
\begin{equation*}\label{Restimate}
\hR\triangleq\R+\Ep\;\Longrightarrow\;\widehat{R}_{ij}=R_{ij}+\ep_{ij}, \; \forall i,j\in[N],
\end{equation*}
where $\{\ep_{ij}\}$ denote the errors in estimating $\{R_{ij}\}$. Hence, replacing $R_{ij}$ with $\widehat{R}_{ij}$ in \eqref{Log_equation} yields\vspace{-0.4cm}
\begin{align*}
\log\left(\widehat{R}_{ij}\right)&\triangleq\widehat{\mu}_{ij}+\jmath\cdot\widehat{\nu}_{ij}=\log\left(R_{ij}\right)+\overbrace{\log\left(1+\frac{\ep_{ij}}{R_{ij}}\right)}^{\triangleq\zeta_{ij}}\\%\label{log_equation_exact}\\
&\triangleq(\mu_{ij}+\varepsilon_{ij})+\jmath\cdot(\nu_{ij}+\epsilon_{ij}),%\label{log_equation_exact2}
\end{align*}
for all $i,j\in[N]$, where $\zeta_{ij}$ is the transformed complex-valued ``measurement noise", with $\varepsilon_{ij}$ and $\epsilon_{ij}$ as its real and imaginary parts, resp. Consequently, we now have the following \emph{exact} linear relations
\begin{equation}\label{exact_linear_equation}
\begin{cases}
\widehat{\mu}_{ij}=\mu_{ij}+\varepsilon_{ij}\\
\widehat{\nu}_{ij}=\nu_{ij}+\epsilon_{ij}\\
\end{cases} \Longrightarrow \; \uy_{\xi}\triangleq\uy+\uxi=\H\utheta+\uxi\in\Rset^{N^2\times 1},
\end{equation}
where $\uy_{\xi}$ is defined exactly as $\uy$, only with $\{\widehat{\mu}_{ij},\widehat{\nu}_{ij}\}$ replacing $\{\mu_{ij},\nu_{ij}\}$, and the ``noise" vector $\uxi\triangleq\left[\uvarep^{\tps} \;\, \uepsilon^{\tps}\right]^{\tps}$ consists of the respective $0.5N(N+1)$ elements of the $\widehat{\mu}_{ij}$-related noise $\varepsilon_{ij}$ in $\uvarep$, and of the $0.5N(N-1)$ elements of the $\widehat{\nu}_{ij}$-related noise $\epsilon_{ij}$ in $\uepsilon$.

Despite a previous claim in the literature that ``the estimation of phase errors is independent of gain errors" (\hspace{1sp}\cite{li2006theoretical}, Subsection III-B), we have recently shown that although the ``noiseless" equations \eqref{real_log_eq}--\eqref{imag_log_eq} are deterministically decoupled, the ``noisy" equations (left-hand side of \eqref{exact_linear_equation}) are in fact \textit{statistically} coupled. That is, the noise terms $\{\varepsilon_{ij}\}$ and $\{\epsilon_{ij}\}$ are correlated, meaning that more accurate estimates would be obtained by jointly estimating all the unknowns, including $\tupsi,\uphi$, together, via a widely linear estimate (e.g., \cite{picinbono1995widely}) based on all the complex measurements $\{\widehat{R}_{ij}\}$ (forming $\uy_{\xi}$), and a unified, full analysis of the transformed noise $\{\zeta_{ij}\}$.

From the celebrated Gauss-Markov theorem \cite{lehmann2006theory}, the Best Linear Unbiased Estimate (BLUE) of $\utheta$ given $\uy_{\xi}$ is the OWLS estimate
\begin{equation*}\label{OWLSestimate}
\widehat{\utheta}_{\tiny{\text{OWLS}}}\triangleq\left(\H^{\tps}\mLambda_{\xi}^{-1}\H\right)^{-1}\H^{\tps}\mLambda_{\xi}^{-1}\left(\uy_{\xi}-\ueta_{\xi}\right),
\end{equation*}
where $\ueta_{\xi}\triangleq\Eset\left[\uxi\right]$ and $\mLambda_{\xi}\triangleq\Eset\left[\left(\uxi-\ueta_{\xi}\right)\left(\uxi-\ueta_{\xi}\right)^{\tps}\right]$ are the mean and covariance matrix of $\uxi$, resp. The BLUE attains the minimal attainable MSE matrix among all linear unbiased estimates. When $\uxi$ is Gaussian (which is true asymptotically in our case from the MLE properties \cite{lehmann2006theory}), $\widehat{\utheta}_{\tiny{\text{OWLS}}}$ is also the MLE of $\utheta$ based on $\uy_{\xi}$, which is efficient \cite{ra1922mathematical}, hence it is also the uniformly minimum-variance unbiased estimate (\hspace{1sp}\cite{lehmann2006theory}) of $\utheta$ based on $\uy_{\xi}$.

\begin{figure}[t]
	\centering
	\begin{subfigure}[b]{0.45\textwidth}
		\includegraphics[width=\textwidth]{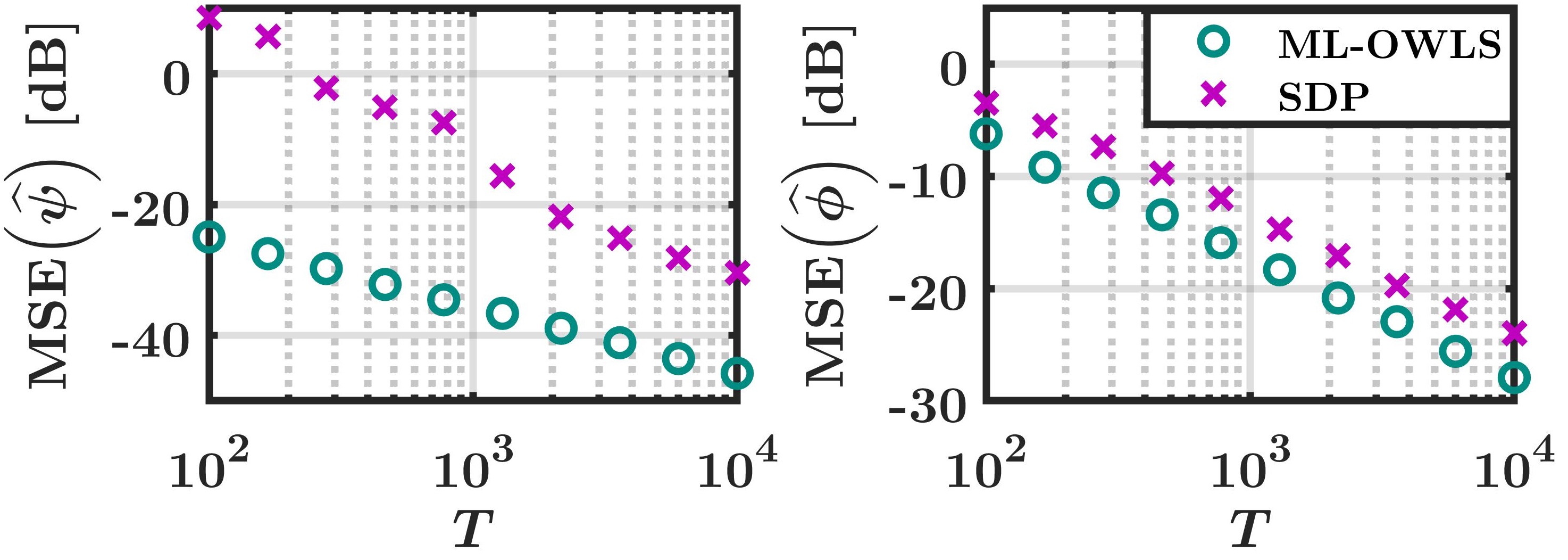}
		\vspace{-0.4cm}
		\caption{}
		\label{fig:MSE_vs_T}
	\end{subfigure}
	\\
	\begin{subfigure}[b]{0.45\textwidth}
		\includegraphics[width=\textwidth]{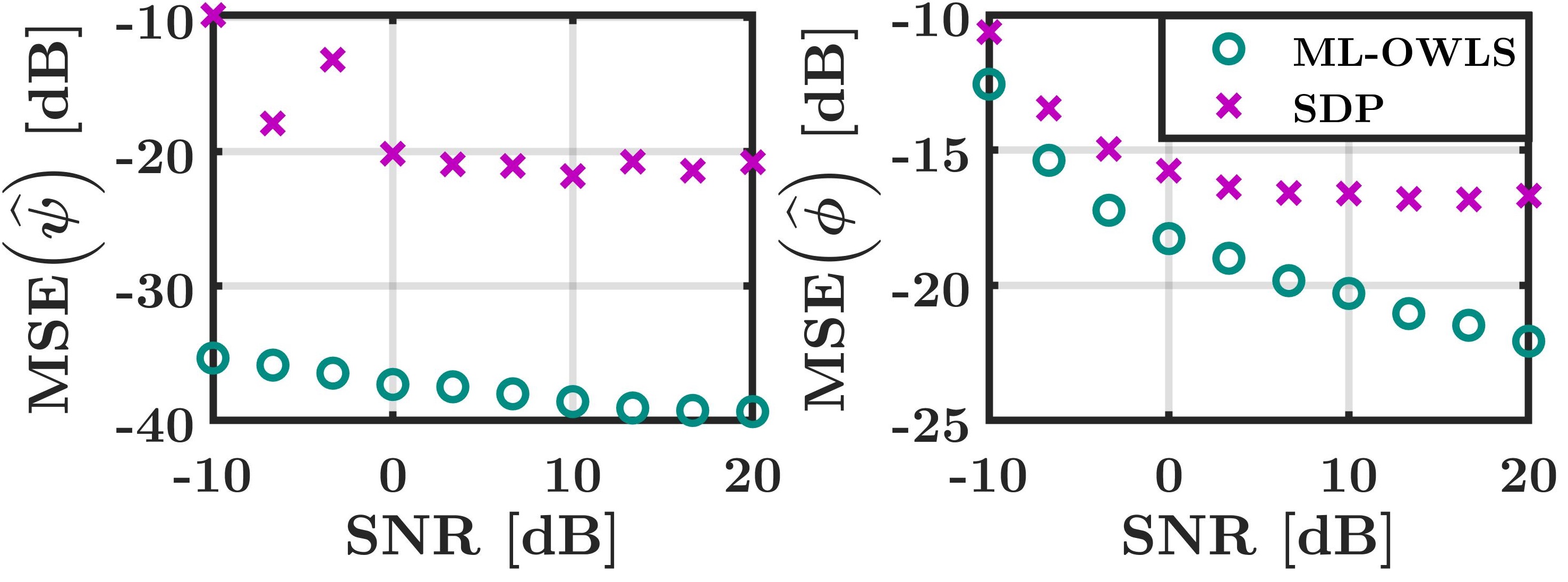}
		\vspace{-0.4cm}
		\caption{}
		\label{fig:MSE_vs_SNR}
	\end{subfigure}
	\vspace{-0.2cm}
	\caption{Average MSEs of $\widehat{\upsi}$ and $\widehat{\uphi}$ (a) vs.\ $T$, at $\text{SNR}\hspace{-0.05cm}=\hspace{-0.05cm}10$[dB], and (b) vs.\ SNR, with $T\hspace{-0.05cm}=\hspace{-0.05cm}2\cdot10^3$. The considerable performance improvement is evident.}
	\label{fig:MSE}
	\vspace{-0.4cm}
\end{figure}

In \cite{weiss2020asymptotically}, addressing blind calibration of ULAs, we showed that 
\begin{equation}\label{noise_approx_mean}
\ueta_{\xi}\approx-\frac{1}{T}\cdot0.5\cdot\left[\uon^{\tps}_{0.5N(N+1)}\; \uo^{\tps}_{0.5N(N-1)}\right]^{\tps}\triangleq\widehat{\ueta}_{\xi},
\end{equation}
for a sufficiently large $T$, such that $|\ep_{ij}|\ll |R_{ij}|$ for all possible $(i,j)$. As for the covariance matrix of $\uxi$, which also reads $\mLambda_{\xi}=\Eset\left[\uxi\uxi^{\tps}\right]-\ueta_{\xi}\ueta_{\xi}^{\tps}$, we have further shown in \cite{weiss2020asymptotically} that for all $i,j,k,\ell\in[N]$,
\begin{align}\label{approximated_covariance_noise}
&\Eset\left[\varepsilon_{ij}\cdot\varepsilon_{k\ell}\right]\approx\frac{1}{T}\cdot0.5\cdot\Re\left\{\frac{R_{ik}R^*_{j\ell}}{R_{ij}R^*_{k\ell}}+\frac{R_{i\ell}R^*_{jk}}{R_{ij}R_{k\ell}}\right\},\\
&\Eset\left[\epsilon_{ij}\cdot\epsilon_{k\ell}\right]\approx\frac{1}{T}\cdot0.5\cdot\Re\left\{\frac{R_{ik}R^*_{j\ell}}{R_{ij}R^*_{k\ell}}-\frac{R_{i\ell}R^*_{jk}}{R_{ij}R_{k\ell}}\right\},\\
&\Eset\left[\varepsilon_{ij}\cdot\epsilon_{k\ell}\right]\approx\frac{1}{T}\cdot0.5\cdot\Im\left\{\frac{R_{i\ell}R^*_{jk}}{R_{ij}R_{k\ell}}-\frac{R_{ik}R^*_{j\ell}}{R_{ij}R^*_{k\ell}}\right\},\label{approximated_covariance_noise_last}
\end{align}
under the same asymptotic regime, so that $\mLambda_{\xi}$ approximately depends on $\R$ \textit{only}. For the full derivation of \eqref{noise_approx_mean}--\eqref{approximated_covariance_noise_last}, see \cite{weiss2020asymptotically}, Appendix B. 

\noindent\textit{\underline{Remarks}:}
%\vspace{-0.05cm}
\begin{enumerate}
	\itemsep0.05em 
	\item[\circled{1}] For a large enough $T$, \eqref{noise_approx_mean}--\eqref{approximated_covariance_noise_last} hold for any fixed SNR; and
	\item[\circled{2}] It is evident from \eqref{approximated_covariance_noise_last} that $\varepsilon_{ij},\epsilon_{k\ell}$ are generally dependent.
    %The correlation \eqref{approximated_covariance_noise_last} shows $\varepsilon_{ij},\epsilon_{k\ell}$ are generally dependent.
\end{enumerate}

Of course, the true $\R$ is in fact unknown. However, since $\hR$ is the MLE of $\R$, by virtue of the invariance property of the MLE \cite{mukhopadhyay2000probability}, it follows that $\hmLambda_{\xi}$, a matrix whose elements are computed by \eqref{noise_approx_mean}--\eqref{approximated_covariance_noise_last}, but with $\{\widehat{R}_{ij}\}$ replacing $\{R_{ij}\}$, is approximately the MLE of $\mLambda_{\xi}$. Therefore, we propose the ``ML-based OWLS" estimate
\begin{gather}\label{MLbasedWLSestimate}
\widehat{\utheta}_{\tiny{\text{ML-OWLS}}}\triangleq\left(\H^{\tps}\hmLambda_{\xi}^{-1}\H\right)^{-1}\H^{\tps}\hmLambda_{\xi}^{-1}\left(\uy_{\xi}-\widehat{\ueta}_{\xi}\right),
\end{gather}
from which the ML-based OWLS estimates of the gains and phases
\begin{align*}%\label{OWLS_estimates}
&\left(\widehat{\psi}_n\right)_{\tiny{\text{ML-OWLS}}}=\exp\bigg(\left(\widehat{\theta}_n\right)_{\tiny{\text{ML-OWLS}}}\bigg), \; \forall n\in[N-1], \\%\label{OWLS_estimates1}\\
&\left(\widehat{\phi}_n\right)_{\tiny{\text{ML-OWLS}}}=\left(\widehat{\theta}_{N-1+n}\right)_{\tiny{\text{ML-OWLS}}}, \; \forall n\in[N-2], %\label{OWLS_estimates2}
\end{align*}
are readily extracted (recall $\tpsi_1=0 \Rightarrow \psi_1=1$ and $\phi_1=\phi_2=0$). Invoking the continuous mapping theorem \cite{mann1943stochastic} and the consistency of the MLE \cite{cramer2016mathematical} $\hmLambda_{\xi}$, for a sufficiently large $T$: $\widehat{\utheta}_{\tiny{\text{ML-OWLS}}}\approx\widehat{\utheta}_{\tiny{\text{OWLS}}}$.

The potential phase wrapping problem---possibly corrupting the \emph{linear} relation \eqref{imag_log_eq} via a modulo operation---can be mitigated using the same method described\footnote{\label{lackspace}Omitted from this paper due to lack of space.} in \cite{weiss2020asymptoticallyicassp}, Subsection 3.2.% The technical description, which follows exactly the same guidelines as in \cite{weiss2020asymptoticallyicassp}, is omitted due to space limitation and will appear in a future work.

\begin{figure}[t]
	\centering
	\includegraphics[width=0.45\textwidth]{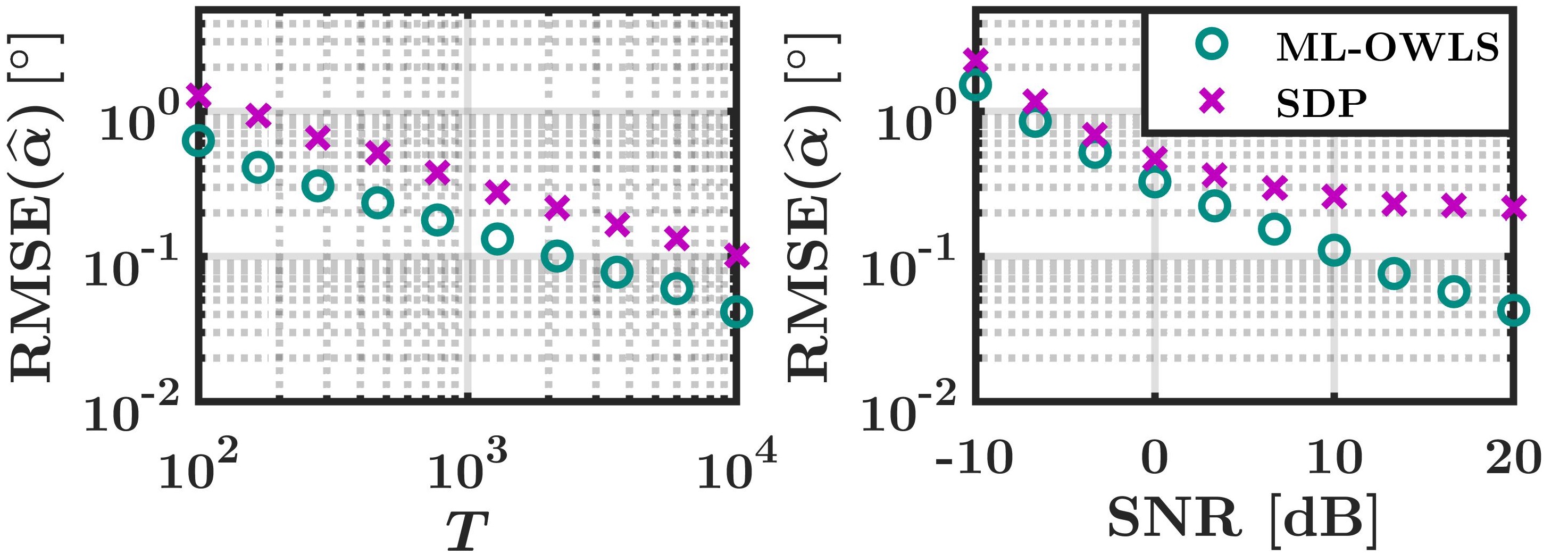}
	\vspace{-0.1cm}
	\caption{DOAs estimates average RMSEs for the post-calibration SS-MUSIC estimates (a) vs.\ $T$, at $\text{SNR}=10$[dB], and (b) vs.\ SNR, with $T=2000$. Clearly, the enhanced calibration enables higher accuracy in DOA estimation.}
	\label{fig:RMSE_DOA}
	\vspace{-0.3cm}
\end{figure}

\vspace{-0.25cm}
\section{Simulation Results}\label{sec:results}
\vspace{-0.15cm}
We consider model \eqref{modelequation} with $N_1=N_2=L=4$, i.e., $N=8$ sensors, $d=\lambda/2$, and $M=15$ unit-variance sources impinging from angles $\ualpha$ located uniformly on $[20^\circ,70^\circ]$. The sensors' gains and phases were fixed to $\upsi=[1\;1.3\;1.1\;0.7\;2.2\;0.9\;1.2\;0.8]^{\tps}$ and $\uphi=[0^\circ\;0^\circ\;5^\circ\;11^\circ-\hspace{-0.05cm}8^\circ\;3^\circ-\hspace{-0.05cm}7^\circ\;9^\circ]^{\tps}$, resp., with $\psi_1,\phi_1$ and $\phi_2$ as known references (w.l.o.g.). Empirical results were obtained by averaging $10^3$ independent trials.

\begin{table}[t]
	\begin{center}
		\begin{tabular}{|c|c|c|}
			\hline
			\multicolumn{2}{|c|}{Average Runtime, $N=8$} \\
			\hline
			ML-OWLS & SDP  \\
			\hline
			{\color{MLOWLSgreen}$0.0032$}[sec] & {\color{SDPpurple}$1.616$}[sec] \\
			\hline
		\end{tabular}
	\end{center}
	\vspace{-0.4cm}
	\caption{Average runtime of the two methods for $N\hspace{-0.05cm}=\hspace{-0.05cm}8$. Our method achieves higher accuracy, while reducing execution time by orders of magnitude.}
	\label{table:Avgruntime}
	\vspace{-0.5cm}
\end{table}

Fig.\ \ref{fig:MSE} presents the average MSEs obtained by $\widehat{\upsi}_{\tiny{\text{ML-OWLS}}}$ and $\widehat{\uphi}_{\tiny{\text{ML-OWLS}}}$. For comparison, we also show the MSEs obtained by Ramamohan \textit{et al.}'s Semidefinite Programming-based (SDP) estimates \cite{ramamohan2019blind}, which were demonstrated as superior to the STLS method \cite{han2015calibrating}. Evidently, our proposed estimates provide considerably improved calibration relative to SDP. Table \ref{table:Avgruntime} shows the average runtime (in [sec], on an Intel i7 CPU at 2.7[GHz]) in this setup for the two calibration methods under consideration. Since the core computation of our method boils down to a closed-form (non-iterative) formula \eqref{MLbasedWLSestimate}, it is substantially more computationally efficient. 

Next, we turn to evaluate the resulting accuracy in DOAs estimation. We consider the same setup, only now with $M=3$ sources, $\ualpha=[33^\circ\,45^\circ\,57^\circ]^{\tps}$, and we assume $M$ is known (or, was successfully detected \cite{han2013improved,garg2020source,weiss2019blind}). Fig.\ \ref{fig:RMSE_DOA} presents the average Root MSE (RMSE) of all the DOAs estimates $\{\widehat{\alpha}_m\}_{m=1}^3$, obtained using Spatial Smoothing MUSIC (SS-MUSIC, e.g., \cite{liu2015remarks}) vs.\ $T$, for fixed $\text{SNR}=10$[dB], and vs.\ the SNR, for fixed $T=2\cdot10^3$. As expected, due to the calibration improvement, the accuracy in DOAs estimation is improved as well.
\vspace{-0.2cm}
\section{Conclusion}\label{sec:conclusion}
\vspace{-0.2cm}
We presented a blind calibration scheme for $2$-level nested arrays. Based on the partially-Toeplitz structure of the observations' nominal covariance matrix, and building upon our recent results for statistically enhanced blind calibration of ULAs, we derived the ML-OWLS estimates of the gain and phase offsets, which were shown to be superior, both in terms of estimation accuracy and of computational complexity, to the recently proposed SDP-based method. Our improved calibration leads to enhanced accuracy in DOAs estimation, which was also demonstrated in simulations.

%We note in passing that it may be shown analytically\footnote{Omitted from this paper due to lack of space.} that, under reasonable conditions, our proposed estimates approximately coincide with the MLEs of the sensors' gain and phase offsets in joint ML estimation of all the unknown model parameters. Furthermore, the proposed calibration scheme may also be used for non-Gaussian signals and still achieve a significant performance improvement (with analytical guarantees).

We note in passing that the proposed calibration scheme may also be used for non-Gaussian signals and still achieve a significant performance improvement with analytical guarantees$^{{\text{3}}}$. Moreover, the concepts presented here for $2$-level nesting can be extended to $K$-levels, $K>2$. These extensions will be presented in future work.

\vspace{-0.2cm}
\section{Acknowledgment}\label{sec:acknowledgment}
\vspace{-0.2cm}
The authors thank Prof.\ Sundeep Prabhakar Chepuri for providing a helpful MATLAB code for the blind calibration algorithm in \cite{ramamohan2019blind}.
\vspace{-0.2cm}
%\appendix
%\vspace{-0.45cm}
%\section{Computation of the Noise Covariance $\mLambda_{\xi}$}\label{AppendxA}
%\vspace{-0.3cm}

%\unappendix
\bibliographystyle{IEEEbib}
\small{\bibliography{refs}}

\end{document}